\documentclass[oldversion]{aa}
\pdfoutput=1

\usepackage[intlimits]{amsmath}
\usepackage{txfonts}
\usepackage[american]{babel}
\usepackage{graphicx}
\usepackage{upgreek}
\usepackage{xspace}

\usepackage{natbib}
\newcommand{\Halpha}{H$\upalpha$\xspace}
\newcommand{\Alfven}{Alfv\'{e}n\xspace}
\newcommand{\Alfvenic}{Alfv\'{e}nic\xspace}
\newcommand{\Rjetn}{\ensuremath{R_{\jet,\nozzle}}}
\newcommand{\taug}{\ensuremath{\tau_\mathrm{g}}}
\newcommand{\tauc}{\ensuremath{\tau_\mathrm{c}}}
\newcommand{\tauK}{\ensuremath{\tau_K}}
\newcommand{\nozzle}{\ensuremath{\mathrm{n}}}
\newcommand{\jet}{\ensuremath{\mathrm{jet}}}

\newcommand{\simm}{\mathord{\sim}}
\newcommand{\proptoo}{\mathord{\propto}}
\newcommand{\Nabla}{\vec{\nabla}}

\newcommand{\abs}[1]{\ensuremath{\left|#1\right|}}
\newcommand{\pmag}{\ensuremath{p_\text{mag}}}

\newcommand{\vA}{\ensuremath{v_\mathrm{A}}}
\newcommand{\vAn}{\ensuremath{v_\mathrm{A,\nozzle}}}

\newcommand{\vArn}{\ensuremath{v_{\mathrm{A}r\mathrm{,n}}}}
\newcommand{\vAphi}{\ensuremath{v_{\mathrm{A}\varphi}}}

\newcommand{\cs}{\ensuremath{c_\mathrm{s}}}
\newcommand{\csn}{\ensuremath{c_\mathrm{s,n}}}
\newcommand{\cso}{\ensuremath{c_{\mathrm{s}0}}}
\newcommand{\er}{\ensuremath{\hat{\vec{e}}_r}}
\newcommand{\ephi}{\ensuremath{\hat{\vec{e}}_\phi}}
\newcommand{\evarphi}{\ensuremath{\hat{\vec{e}}_\varphi}}

\newcommand{\de}{\ensuremath{\mathrm{d}}}
\newcommand{\const}{\ensuremath{\text{const}}}
\newcommand{\degree}{\ensuremath{^\circ}}
\newcommand{\efrate}{\ensuremath{\mathcal{E}}}

\begin{document}

\title{Kink instabilities in jets from rotating magnetic fields}
\author{R. Moll \and H. C. Spruit \and M. Obergaulinger}
\offprints{\protect\raggedright R. Moll,\\\email{rmo@mpa-garching.mpg.de}}
\institute{Max-Planck-Institut f\"{u}r Astrophysik, Karl-Schwarzschild-Str. 1, 85741 Garching, Germany}
\date{Accepted on September 15, 2008}
\abstract{We have performed 2.5D and 3D simulations of conical jets driven by
the rotation of an ordered, large-scale magnetic field in a stratified
atmosphere.  The simulations cover about three orders of magnitude in distance
to capture the centrifugal acceleration as well as the evolution past the
\Alfven surface.  We find that the jets develop kink instabilities, the
characteristics of which depend on the velocity profile imposed at the base of
the flow. The instabilities are especially pronounced with a rigid rotation
profile, which induces a shearless magnetic field. The jet's expansion appears
to be limiting the  growth of \Alfven mode instabilities.
}
\maketitle

\section{Introduction}

Strong magnetic fields on large scales may play an essential, active role in
the formation and evolution of jet-like outflows.  The general idea is that a
poloidal magnetic field, embedded in a plasma and anchored e.g. in an accretion
disk or a black hole, is forced into rotation at the anchor point, a toroidal
field develops and the plasma is accelerated by what can be interpreted as a
centrifugal force in a corotating frame
\citep{1982Blandford,1996Spruit,2000Koenigl}.  However, this magnetocentrifugal
acceleration is only effective up until the \Alfven surface, defined as the
surface where the flow velocity equals the \Alfven velocity.  Beyond this
point, the magnetic field will be strongly wound-up. Such a field configuration
is potentially unstable with respect to certain MHD instabilities.

MHD jets are susceptible to a variety of instabilities.  Kelvin-Helmholtz (KH)
instabilities are fed by the relative kinetic energy between the jet and the
ambient medium.  They can distort the jet surface only (ordinary modes) or the
whole beam \citep[e.g.][]{1991Birkinshaw}, provoking shocks, mixing with
ambient material and possibly a disruption of the jet
\citep{1995Bodo,1998Bodo}.  The presence of strong magnetic fields, poloidal or
toroidal, is expected to hamper the growth of KH instabilities
\citep{1992Appl,1999Keppens}.

The free energy associated with the toroidal magnetic field is responsible for
another class of instabilities, which is traditionally known as current-driven
(CD) and of notorious importance in controlled fusion devices \citep[for an
introduction, see, e.g.][]{1987Freidberg,1978Bateman}.  The relevance for
magnetized astrophysical jets has been pointed out by
\citet{1993Eichler,1997Spruit,1998Begelman} and others.  Among CD
instabilities, $m=1$ kink instabilities are the most effective.  An ideal kink
mode is characterized by helical displacements of the cylindrical cross
sections of a plasma column.  It is expected to grow on a dynamical time scale
with respect to an \Alfven wave crossing the unstable column. The
susceptibility is strongly dependent on the magnetic pitch, a measure for the
degree of wind-up.  Kink instabilities might destroy the ordered, symmetric
state of a jet, leading to its disruption or, through magnetic reconnection,
the associated dissipation of magnetic fields and steepening of the magnetic
pressure gradient, to its acceleration \citep{2002Drenkhahn,2006Giannios}.

Different kinds of instability can mix and interact. For example,
\citet{2002Baty} show how CD instabilities can stabilize KH vortices at the jet
surface. For this work, we used conditions under which CD kink instabilities
are expected to dominate (low plasma-$\beta$, small magnetic pitch).

For a self-consistent study of kink instabilities, numerical simulations need
to be carried out in 3D.  \citet{1996Lucek} did so using a simple model in
which a toroidal magnetic field configuration was allowed to expand into a
uniform atmosphere. This generated a jet which was subject to kink
instabilities. \citet{2004Nakamura} performed 3D simulations of MHD jets in
variously stratified atmospheres, finding that they can develop kink-like
distortions in the trans-\Alfvenic region.  Laboratory experiments of MHD jets
have been performed by \citet{2005Hsu}, confirming that the magnetic pitch
plays a crucial role for the formation of kink instabilities.

\subsection{Effects of jet expansion}

Jets from protostars, and especially AGN and microquasars, expand in width $d$
by orders of magnitude after passing through their \Alfven radius. In an
expanding flow there is no clean separation between time dependence due to
instability and that due to the expansion itself, making the question of
stability less well defined. Analytical studies thus tend to focus on
instabilities in a cylindrical geometry, with constant diameter (such as in the
``magnetic tower'' picture of \citealt{2003Lynden}). Expansion has strong
consequences on the behavior of instabilities, however, compared with jets
modeled as cylinders of constant width.

First, there is the tendency for the toroidal (azimuthal, around the jet axis)
component of the magnetic field to dominate in an expanding jet. From the
induction equation, the poloidal and toroidal  components of the field vary as
$B_\mathrm{p}\sim 1/d^2$ and $B_\varphi\sim 1/d$ respectively (for constant
jet velocity). Expansion thus causes a continual increase of the ratio
$B_\varphi/B_\mathrm{p}$. Even when dissipation were to decrease the toroidal
field at some point, the ratio increases again on further expansion. Free
energy available in the toroidal field thus remains the dominant form of
magnetic energy, and one may expect the question of stability and dissipation
to remain relevant on all length scales. It also follows that the nonlinear
development of instabilities in an expending jet is expected to be very
different from the constant-diameter case.

Secondly, expansion has a strong effect even on the conditions for occurrence
of instability. It has a stabilizing effect, since magnetic instabilities
become ineffective when their signal speed (the \Alfven speed) drops below the
lateral expansion speed of the jet. This is discussed further below.

\subsection{Rationale of the calculations}

The aim of the calculations reported here is to study how kink instability
operates under these conditions of expansion of the jet over several orders of
magnitude in width.

The degree of instability to be expected in a jet driven by a rotating magnetic
field is intimately tied to the way it is collimated.  If, instead of being
cylindrical, the jet has a non-vanishing opening angle $\vartheta$, the
expected incidence of instability depends on the details of the
dependence of $\vartheta$ on distance $r$.  An opening angle increasing
with distance reduces instability, while for an asymptotically vanishing
opening angle instability must always set in at some distance, if it was not
present already from the start (see discussion in Sect.~\ref{sec:growthexp}).

The setup in the simulations presented here produces jets in the intermediate
case of an (approximately)  constant opening angle: a ``conical'' outflow. It
turns out that in this case the presence of instabilities and their amplitude
depends on secondary conditions such as the rotation profile imposed at the
base of the flow, hence it is a good test case for the incidence of
instabilities.

Since the observed jets travel over such large distances, even marginal forms
of instability can become effective. An important goal of the present
calculations is therefore to cover a large range in distance, about 3 orders of
magnitude. This is achieved by the use of a grid adapted to the approximately
conical shape of the jet.

\subsection{Expected instability growth in expanding jets}
\label{sec:growthexp}

In the following, we estimate how the sideways expansion affects the growth of
instabilities in broadening jets. In spherical coordinates
$(r,\vartheta,\varphi)$, we assume that the jet radius (distance to the jet's
central axis) is given by
\begin{equation}
    R = R' \left( \frac{r}{r'} \right)^\alpha \quad \text{with} \quad R' = r' \sin\vartheta'
\label{}
\end{equation}
where the prime stands for evaluation at a reference distance $r'$, for which
we take a distance somewhat beyond the \Alfven radius. The flow has then
approximately reached its asymptotic speed $v_r \approx \const$, and the
magnetic field has become predominantly azimuthal. In the absence of magnetic
dissipation due to instabilities, the field strength then varies as $B_\varphi
\propto R^{-1}$ (magnetic flux conservation) and the density as $\rho \propto
R^{-2}$ (mass conservation). Since the growth rate $\varGamma_\mathrm{g}$ is
expected to scale with the \Alfven crossing rate $\vAphi/R$, we introduce a
dimensionless instability rate $\varkappa$ of order unity:
\begin{equation}
    \varGamma_\mathrm{g} = \varkappa \frac{\vAphi}{R}
        = \varkappa \frac{\vAphi}{R'} \left( \frac{r}{r'} \right)^{-\alpha}
\label{}
\end{equation}
where $\vAphi=B_\varphi/\sqrt{4\pi \rho}$ is the azimuthal \Alfven velocity.
The expansion rate is estimated by
\begin{equation}
    \varGamma_\mathrm{e} = \frac{\de \ln R}{\de t}
                      = \frac{1}{R} \frac{\de r}{\de t} \frac{\de R}{\de r} = \frac{\alpha v_r}{r} .
\label{}
\end{equation}
We find
\begin{equation}
    \frac{\varGamma_\mathrm{g}}{\varGamma_\mathrm{e}} =
            \frac{\varkappa}{\upsilon \alpha \sin \vartheta'} \left( \frac{r}{r'} \right)^{1-\alpha}
\label{}
\end{equation}
with $\upsilon \coloneqq v_r/\vAphi \approx \const$ according to the ballistic
approximations mentioned above.  Consequently, the instability growth rate
dominates at some distance $r$ for a collimating jet ($\alpha<1$).
Decollimation ($\alpha>1$), on the other hand, thwarts the growth of
instabilities. A conical jet ($\alpha = 1$) constitutes a limiting case where
all depends on the combination of parameters $\varkappa/(\upsilon\, \sin
\vartheta')$, which is of order unity.  A numerical simulation is necessary to
find out whether the instability or expansion prevails.

The paper is organized as follows. In Sect.~\ref{sec:model}, we introduce the
magnetocentrifugal jet model and account for the assumptions made in our
simulations.  A detailed description of the numerical setup, the coordinate
system and the scale-free units employed in the analysis is given in
Sect.~\ref{sec:methods}.  In Sect.~\ref{sec:simulatedcases} we give the
parameters of the simulated cases and in Sect.~\ref{sec:results} we present the
results.  There, we start by making predictions on the characteristics of
instabilities by examining the relevant properties of our simulated jets. We
proceed with an analysis of the instabilities that actually appeared and
complete with looking for effects on the jets' dynamics. We finish with a
discussion and conclusions in Sect.~\ref{sec:discussion}.

\section{The model}
\label{sec:model}

The model is  construed to apply to jets produced by ordered magnetic fields
anchored in an accretion disk. This has become the default interpretation for
the jets observed in AGN, microquasars and protostellar objects, though it must
be kept in mind that observational evidence of the key ingredient in this
model, the ordered field \citep{1976Bisnovatyi,1982Blandford}, is still somewhat
indirect. 

More uncertain is the shape of this field. The strength of the field anchored
in the disk is likely to scale in some way with the orbital kinetic energy (or
gas pressure) in the disk, hence will decline with distance $R$ from the
rotation axis. In the absence of more detailed information, we consider a
simple form for a field of this kind, one in which the vertical (normal to the
disk) component at the surface $B_z$ varies as $B_z(R) \propto
[1+(R/z_0)^2]^{-\nu}$.  Neglecting gas pressure and fluid motions, the field
above the disk would be a potential field, its shape defined uniquely by $B_z$.
For $\nu=3/2$ it is the field of a monopole with the source at a depth $z_0$
below the center of the disk.

The initial state of the model is a gas distribution in hydrostatic equilibrium
in a field of this monopolar shape. Rotation is applied at the lower boundary,
in a region $R<R_0$ (see Sect.~\ref{sec:iandbcs} for details). This generates
an outflow with an approximately constant opening angle on the order $R_0/z_0$
(a ``conical'' outflow). The surrounding volume remains approximately in static
equilibrium, and serves to collimate the outflow to the desired opening angle.

The magnetic field responds to the rotation by winding up. That is, a toroidal
magnetic field $B_\varphi$ is produced and gives, together with the poloidal
field $B_r$, rise to helical field lines.  The magnetic pressure gradient
(minus the tension force) associated with $B_\varphi$ gives rise to a poloidal
acceleration of the plasma which is of centrifugal nature in a corotating
frame.  Beyond the \Alfven radius, the acceleration ceases to be effective,
while the field becomes predominantly azimuthal.  The further development
depends on how strongly the jet is affected by instabilities in this highly
wound-up field.  Possibly, they endanger the jet's integrity and/or facilitate
magnetic reconnection events. Magnetic field dissipation can entail further
acceleration of the jet \citep{2002Drenkhahn}.  As discussed above
(Sect.~\ref{sec:growthexp}), the ``conical outflow'' produced in our monopolar
background field is of special interest as it marks the boundary between cases
expected to be strongly respectively weakly unstable. 

A self-consistent investigation of the problem requires a full 3D treatment,
because kink instabilities are non-axisymmetric.  In addition to every 3D
simulation we also performed an axisymmetric (2.5D) simulation with the same
boundary and initial conditions.  This way, we could detect whether the jet
evolves differently due to the instabilities. 

The basic parameters of the model are the magnitude of the rotation velocity,
its profile $\varOmega(R)$, the relative strength of the magnetic field as
measured by plasma-$\beta$ of the initial state, and the jet's opening angle.
The parameter values are chosen such that the \Alfven radius of the resulting
outflow is located within the computational volume, so that the centrifugal
acceleration process is covered in the simulation, but close to the lower
boundary, so that the subsequent evolution can be followed over as large a
distance as numerically feasible. Increasing the imposed rotation rate moves
the \Alfven radius toward the lower boundary.  Due to numerical limitations,
however, $v_\varphi$ could not be increased indefinitely in the simulations,
and a compromise was necessary.  In the results reported below the region
inside the \Alfven radius occupies about 10--20\% of the box length.

\section{Methods}
\label{sec:methods}

We employ a spherical grid for our jet simulations.  This enables us to follow
jets with opening angles over a much longer distance than is possible with a
Cartesian grid, because the jet need not be overresolved at large heights in
order to properly resolve its base.  The obvious choice of letting the jet
propagate along the polar axis is numerically problematic if non-axisymmetric
flows are involved, because the grid is singular there. We therefore let the
jet flow in equatorial direction. The computational volume covers a range
$\Delta\theta=\Delta\phi$ in the polar and azimuthal angles, adjusted to
the opening angle of the flow.

\subsection{MHD equations and numerical solver}
\label{sec:equations}

We numerically solved the ideal adiabatic MHD equations, including a
temperature-dependent temperature-control term $K=K(T(t))$. Explicitly, the
equations are:
\begin{gather}
    \frac{\partial \rho}{\partial t} + \Nabla \cdot (\rho \vec{v}) = 0 , \label{eq:continuity_equation}\\
    \frac{\partial \vec{v}}{\partial t} + \vec{v} \cdot \Nabla \vec{v} = 
        - \frac{1}{\rho} \Nabla p + \frac{1}{4 \pi \rho} (\Nabla \times \vec{B} ) \times \vec{B}
        - \Nabla \Phi,    \label{eq:momentum_equation}\\
    \frac{\partial e}{\partial t} + \Nabla \cdot \left[ \left( e+p+\frac{B^2}{8\pi} \right) \vec{v}
        - \frac{1}{4 \pi} \vec{B}(\vec{B} \cdot \vec{v}) \right]
        = - \rho \vec{v} \cdot \Nabla \Phi + K,   \label{eq:energy_equation} \\
    \frac{\partial \vec{B}}{\partial t} = \Nabla \times (\vec{v} \times \vec{B}), \label{eq:induction_equation}
\end{gather}
where
\begin{equation}
    e = \frac{1}{2} \rho v^2 + \frac{B^2}{8\pi} + \frac{p}{\gamma - 1}
\label{eq:total_energy}
\end{equation}
is the total energy density, $\gamma = 5/3$ is the adiabatic index, $p$ is the
gas pressure, $\Phi$ is the gravitational potential (external, no self-gravity)
and the other symbols have their usual meanings. A notorious problem with
low-$\beta$ MHD simulations in fully conservative form, as in the code used
here, is the amplification of discretization errors that occurs because the gas
pressure is only a small contribution to the  total energy
(cf.~\ref{eq:total_energy}), which is dominated by the magnetic energy. As in
the case of highly supersonic flows, these errors manifest themselves in the
form of ``negative pressures'' at occasional grid points. This problem does not
occur when an equation for the thermal energy equation is used instead of the
total energy. We compute, in parallel, an alternative update of the gas
pressure from the thermal energy equation, in the form
\begin{equation}
    \frac{\partial p}{\partial t} + \Nabla \cdot ( p \vec{v} )
            = - (\gamma-1) p \Nabla \cdot \vec{v} + K.
\label{eq:intenergy_equation}
\end{equation}
Where negative pressures appear, they are replaced by this value.

Another device that turns out very useful to avoid negative pressures is the
temperature-control term $K$ in Eq.~\eqref{eq:energy_equation}. For this we use
a scheme loosely modeled after Newtonian cooling, or an optically thin
radiative loss process. After every full time step, we add/subtract thermal
energy according to \begin{equation} \frac{\Delta p(t)}{p(t=0)} = - \frac{T(t)
- T(t=0)}{T(t=0)} \frac{\Delta t}{\tauK} , \label{eq:cooling_term}
\end{equation} where $T$ is the temperature and $\tauK$ is a time scale chosen
so as to keep the temperature within about a factor 30 of the initial
atmospheric value.

With the MHD Poynting vector
\begin{equation}
    \vec{S} = - \frac{1}{4\pi} ( \vec{v} \times \vec{B} ) \times \vec{B},
\label{}
\end{equation}
Eq.~\eqref{eq:energy_equation} can also be written as
\begin{equation}
    \label{eq:energy_equation_alt}
    \frac{\partial ( e + \rho \Phi) }{\partial t}
       + \Nabla \cdot \left[ 
            \left( \frac{1}{2} \rho v^2 + \frac{\gamma}{\gamma-1}p + \rho \Phi \right) \vec{v}
                           + \vec{S} \right] = K ,
\end{equation}
describing the change of total energy including gravitational potential energy.
We will employ this form later when we look at the energy flow rates.

We used a newly developed Eulerian MHD code (Obergaulinger et al., in
preparation) to solve
Eqs.~(\ref{eq:continuity_equation}--\ref{eq:induction_equation}).  It is based
on a flux-conservative finite-volume formulation of the MHD equations and the
constraint transport scheme to maintain a divergence-free magnetic field
\citep{1998Evans}.  Using high-resolution shock capturing methods
\citep[e.g.,][]{1992LeVeque}, it employs various optional high-order
reconstruction algorithms and approximate Riemann solvers based on the
multi-stage method \citep{2006Toro}.  The simulations presented here were
performed with a fifth order monotonicity-preserving reconstruction scheme
\citep{1997Suresh}, together with the HLL Riemann solver \citep{1983Harten}
and third order Runge-Kutta time stepping.

\subsection{Grid coordinates}

\begin{figure}[t]
\begin{center}
\includegraphics[width=\linewidth]{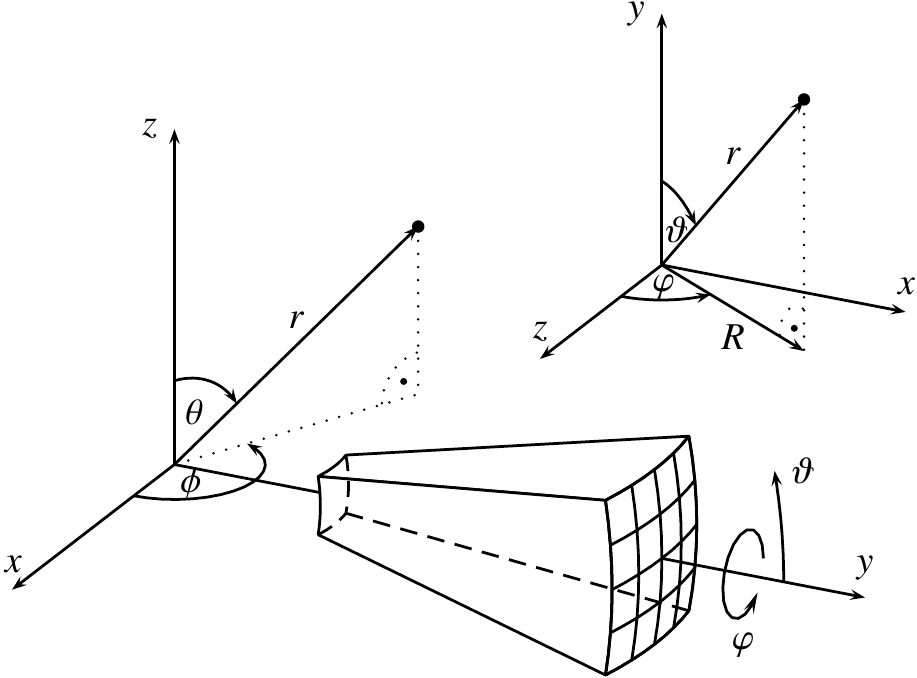}
\end{center}
\caption{Computational domain and coordinate nomenclature (schematic drawing).
The jet propagates along the $y$-axis ($\theta=\phi=\pi/2$), near which the
grid is quasi-Cartesian in the normal plane. To describe the results of the
simulations, we use the alternate spherical coordinate system
$(r,\vartheta,\varphi)$ shown in the upper right inset, which takes the
$y$-axis as the polar axis.}
\label{fig:coords}
\end{figure}

In the 3D simulations, our computational domain was centered around the
$y$-axis in a ``standard'' spherical coordinate system $(r,\theta,\phi)$ with
$\theta$ being the polar angle from the $z$-axis and $\phi$ being the azimuthal
angle about the $z$-axis in the $x$-$y$-plane.  See Fig.~\ref{fig:coords} for
an illustration.  Positioning the domain in equatorial (rather than polar)
direction yields a grid which is free of singularities and quasi-regular in
transverse jet direction: Near the $y$-axis, the distance between two
$\phi=\const$ curves is $\Delta x \approx r\,\Delta\phi$ if we neglect terms of
order $(\theta-\pi/2)^2$ and of order $(\phi-\pi/2)^3$. The distance between
two $\theta=\const$ curves is then $\Delta z \approx r\,\Delta\theta$.  With
uniform spacings  $\Delta\theta$ and $\Delta\phi$, we thus obtain a grid
whose area elements are approximately those of an equidistant Cartesian grid in
a plane normal  to the $y$-axis.

For data analysis and plotting we use the ``alternative'' spherical coordinate
system $(r,\vartheta,\varphi)$ with $\vartheta$ being the polar angle from the
$y$-axis and $\varphi$ being the azimuthal angle about the $y$-axis in the
$z$-$x$-plane. It is adapted to the propagation direction of the jet and as
such better suited to describe its physics.  $R = r\sin\vartheta$ denotes the
distance to the $y$-axis.  Generally, we refer to the $r$-direction with
``poloidal'' or ``radial'', the $\varphi$-direction with ``toroidal'' or
``azimuthal'' and the $y$-axis as ``polar axis'' or ``central axis''.

In the 2.5D simulations, the jet propagates along the $z$-axis and the
$\phi$-direction is taken to be symmetric. However, to avoid confusion, we use
the same nomenclature as in the 3D simulations throughout this paper.  That is,
we visualize the jet as propagating in $y$-direction and employ the
$(r,\vartheta,\varphi)$ system for describing its physics.

For a proper resolution at all radii, it turned out to be necessary to employ
logarithmic grid spacing \citep{1996Park} in $r$-direction.  The $r$-left
interface of grid cell $i$ is situated at
\begin{equation}
    r_\mathrm{l}^i = r_\mathrm{l}^0 \left( \frac{r_\mathrm{r}^{n-1}}{r_\mathrm{l}^0} \right)^{i/n}
\end{equation}
where $n$ is the total number of cells in the domain which is bounded by
$r=r_\mathrm{l}^0$ and $r=r_\mathrm{r}^{n-1}=r_\mathrm{l}^n$. The cell center
of grid cell $i$ is situated at $r_\mathrm{c} =
(r_\mathrm{l}^i+r_\mathrm{r}^i)/2$.

\subsection{Initial and boundary conditions}
\label{sec:iandbcs}

The initial magnetic field is
\begin{equation}
    \vec{B}(r) = \frac{g}{r^2} \er,
\label{}
\end{equation}
corresponding to a magnetic monopole of charge $g$ located at the coordinate
origin.  The associated vector potential
\begin{equation}
    \vec{A} = \frac{g}{r} \tan \left( \frac{\theta}{2} \right) \ephi
\label{}
\end{equation}
was employed in the numerical setup to ensure solenoidality of the discretized
magnetic field.  Satisfying $\Nabla \times \vec{B}=0$, the initial magnetic
field is force-free.

We impose the static gravitational field of a point mass $M$, also located at
the origin. The stratification of gas pressure in this potential is chosen such
that the plasma-$\beta \coloneqq p / \pmag$ in the initital state is constant
throughout the computational domain. Hence, since $B \propto r^{-2}$, $p
\propto r^{-4}$.  The density  in the initial state is determined from
hydrostatic  equilibrium: $\rho GM = - r^2 \de p / \de r \propto r^{-3}$ where
$G$ is the gravitational constant. The
temperature, sound speed and \Alfven velocity vary as $T \propto r^{-1}$, $\cs
\propto r^{-1/2}$ and $\vA \propto r^{-1/2}$ in this stratification.

The lower boundary of the computational volume, where the jet's ``nozzle''
resides, is at a distance $r_\nozzle$ from the origin. The conditions at this
surface are related to the gravitational potential by 
\begin{equation}
    \Phi(r) = -\frac{GM}{r} \quad \text{with} \quad
        M = \frac{4 p_\nozzle r_\nozzle}{G \rho_\nozzle} ,
\label{}
\end{equation}
where the subscript $\nozzle$ (for ``nozzle'') denotes the values at $r_\nozzle$.

At the sides ($\theta$ and $\phi$) and top (upper $r$) of the domain, we use
open boundaries which allow for an almost force-free outflow of material: $p$,
$\rho$, all components of $\vec{v}$ and the transverse components of $\vec{B}$
are mirrored across the boundary interface to the opposing ``ghost cells'', the
normal component of $\vec{B}$ is determined by the solenoidality condition.
The open boundaries work well and cause only minimal artefacts in the form of
reflections.  At the bottom (lower $r$) of the domain, where the jet emanates,
$p$ and $\rho$ are kept fixed at their initial values in all ghost cells.  The
magnetic field is determined by extrapolation from the interior of the domain
using the same scheme as for open boundaries.  The velocity is prescribed to be
zero except for the ghost cells below the nozzle area ($R \le \Rjetn$ at
$r=r_\nozzle$). There, an azimuthal velocity field $\vec{v}=v_\varphi \evarphi$
is maintained, with either a Keplerian velocity profile
\begin{equation}
    v_\varphi = \begin{cases} v_{\varphi,\nozzle}^\text{max} \sqrt{\frac{0.2\Rjetn}{R}}
        & \text{for } 0.2 \Rjetn \le R \le \Rjetn \\
        0
        & \text{elsewhere}
        \end{cases}
\label{eq:Keplerian}
\end{equation}
or a rigid rotation profile
\begin{equation}
    v_\varphi = \begin{cases} v_{\varphi,\nozzle}^\text{max} \frac{R}{\Rjetn}
        & \text{for } R \le \Rjetn \\
        0 & \text{elsewhere} .
        \end{cases}
\label{eq:rigid}
\end{equation}
Note that we use the term ``Keplerian'' to indicate only that $v_\varphi
\propto R^{-1/2}$. The central mass $M$ only serves to balance our chosen
stratification and is not to be understood as the center of an accretion disk.

\subsection{Units}

\begin{table}
\caption{Normalization units}
\centering
\begin{tabular}{ccc}
\hline\hline
Quantity & Symbol(s) & Unit \\
\hline
    length          & $x$,$y$,$z$,$r$,$R$,$h$   & $l_0$ \\
    gas pressure    & $p$                   & $p_0$ \\
    density         & $\rho$                & $\rho_0$ \\
    velocity        & $v$                   & $\cso = \sqrt{ \gamma p_0 / \rho_0 }$ \\
    time            & $t$,$\tau$            & $t_0 = l_0 / \cso$ \\
    energy density  & $e$                   & $p_0$ \\
    energy flow rate& $\efrate$             & $p_0 l_0^3 / t_0$ \\
    force density   & $F$                   & $p_0 / l_0$ \\
    magnetic flux density       & $B$       & $B_0 = \sqrt{8 \pi p_0} $ \\
    current density & $j$                   & $j_0 = B_0 c / (4\pi l_0)$ \\
\hline
\end{tabular}
\label{tab:units}
\end{table}

The setup described above is unambiguously determined by 6 parameters,
$B_\nozzle$,  $p_\nozzle$, $\rho_\nozzle$, $R_{\jet,\nozzle}$, $r_\nozzle$ and
$v_{\varphi,\nozzle}^\text{max}$, but they are not all independent. As units of
length, pressure and density we use $l_0 \equiv 2 R_{\jet,\nozzle}$, $p_0
\equiv p_\nozzle$ and $\rho_0 \equiv \rho_\nozzle$. The physical quantities
expressed in these units are  listed in Table~\ref{tab:units}. Since these
units are arbitrary, the number of independent parameters defining the problem
reduces to $6-3=3$.  These are a plasma-$\beta$ value (which determines
$B_\nozzle$), an opening angle $\vartheta_\mathrm{o} \coloneqq \arcsin ( l_0 /
2 r_\nozzle)$, and a Mach number for the rotation: either
$v_{\varphi,\nozzle}^\text{max} / \csn$ or $v_{\varphi,\nozzle}^\text{max} /
\vAn$. 

The sound speed, \Alfven velocity and escape velocity at $r=r_\nozzle$,
expressed in the normalization units, are $\csn = \cso$,
\begin{align}
    \vAn &= \sqrt{\frac{2}{\gamma}} \frac{B_\nozzle}{B_0} \cso
                        \approx 1.1 \frac{B_\nozzle}{B_0} \cso \\
 \text{and} \quad   v_{\text{esc},\nozzle} &= \sqrt{\frac{8}{\gamma}} \cso \approx 2.2 \cso .
\end{align}

For the sake of clarity, we usually omit the normalization unit.  For example,
$v=5$ would denote a velocity of $5\cso$, which is sonic Mach $5$ at the jet
nozzle.

\section{Cases studied}
\label{sec:simulatedcases}

In the following we present the results of two 3D simulations for two different
rotation profiles imposed at the nozzle, the Keplerian and rigid rotation
profiles given by  (\ref{eq:Keplerian}, case K3) and (\ref{eq:rigid}, case R3).
These are compared with two 2.5D simulations with the same initial and boundary
conditions (cases K2 and R2).

In all cases, the initial state has a constant plasma-$\beta$ of $1/9$
($B_\nozzle=3$), the maximum rotation velocity at the boundary is
$v^\text{max}_{\varphi,\nozzle} = 0.33 \csn = 0.1 \vArn$ and the initial (half)
opening angle is $\vartheta_\mathrm{o} = 5.7\degree$ ($r_\nozzle = 5$).  This
choice of parameters yields a jet with a magnetic pitch low enough to be
unstable to kinks. At the same time, it avoids numerical problems found to
arise with higher (supersonic) rotation velocities as a boundary condition.
The ``grid noise'' in the 3D simulations (the grid is not axisymmetric in the
rotation direction $\varphi$) turned out to be sufficient to excite
instabilities, so we did not need to apply a perturbation by hand. For the
temperature-control term, we used $\tauK=2$.

In the 3D simulations, we used 384 logarithmically spaced grid cells in radial
direction and 96 uniformly spaced grid cells in each of the two angular
directions.  The physical extent of the simulated domain was $500$ in the
radial direction and $33.8\degree$ in each angular direction.  The ratio
between the maximum and minimum $r$ is $505/5 = 101$. The 2.5D simulations were
performed with the same resolution in the radial direction and $64$ grid cells
in the evolved angular direction which had an extent of $16.9\degree$. The 3D
simulations each ran for about one week (wall clock time) on 64 processors with
MPI parallelization.

\section{Results}
\label{sec:results}

\begin{figure}[t]
\begin{center}
\includegraphics[width=\linewidth]{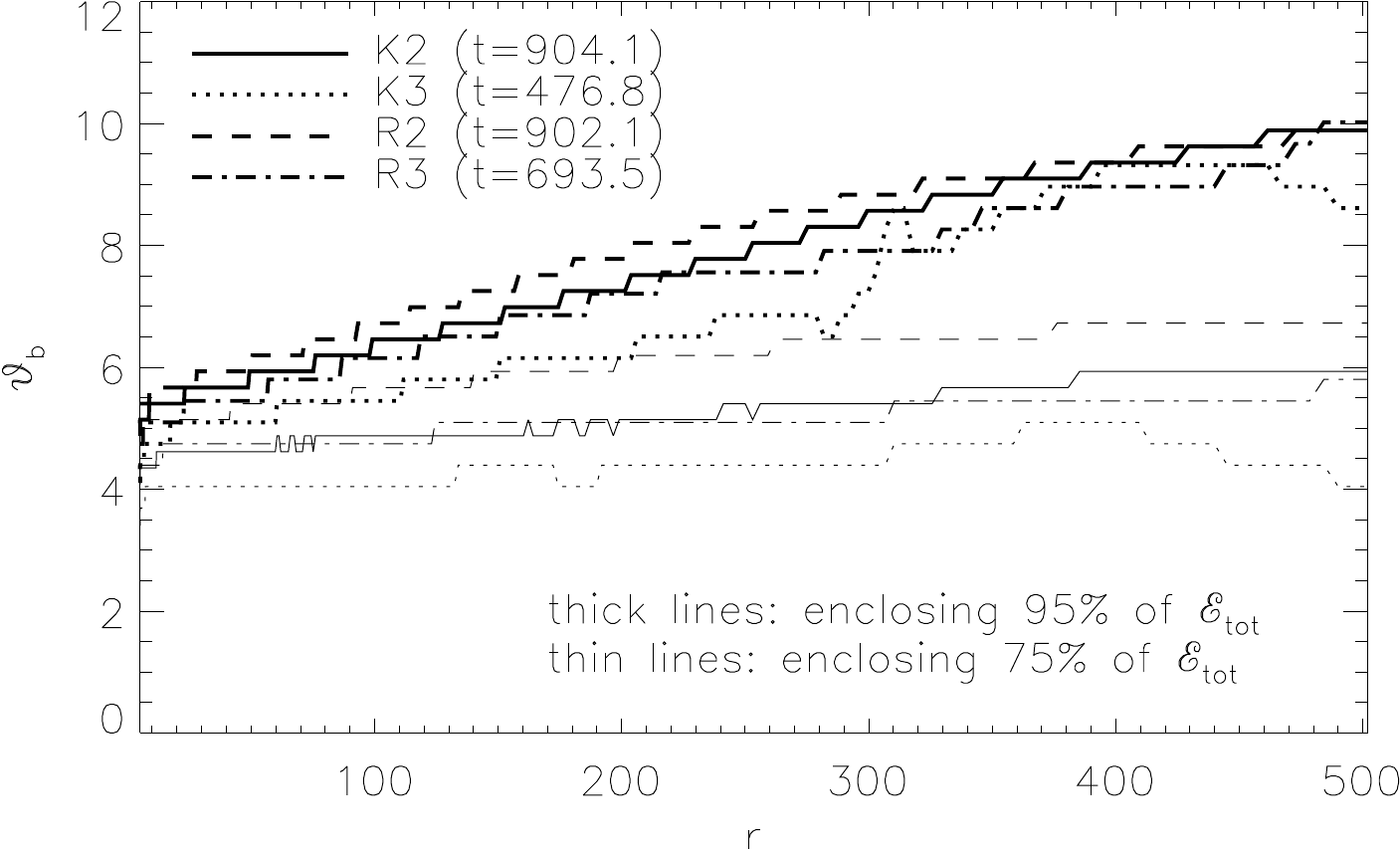}
\end{center}
\caption{Jet border, defined as the polar angle within which a given percentage
of energy flows (see Eq.~\ref{eq:efrate} for the definition of
$\efrate_\text{tot}$), as a function of distance.  In all cases, most of the
energy flow is contained within the
$\vartheta\approx\vartheta_\mathrm{o}=5.7\degree$ surface, where
$\vartheta_\mathrm{o}$ is the initial opening angle.  A minor but increasing
amount of energy flows outside this angle. The energy density in the jet
(especially the toroidal field) causes it to decollimate somewhat compared with
the conical configuration in which it is embedded.}
\label{fig:border}
\end{figure}

\begin{figure}[t]
\begin{center}
\includegraphics[width=\linewidth]{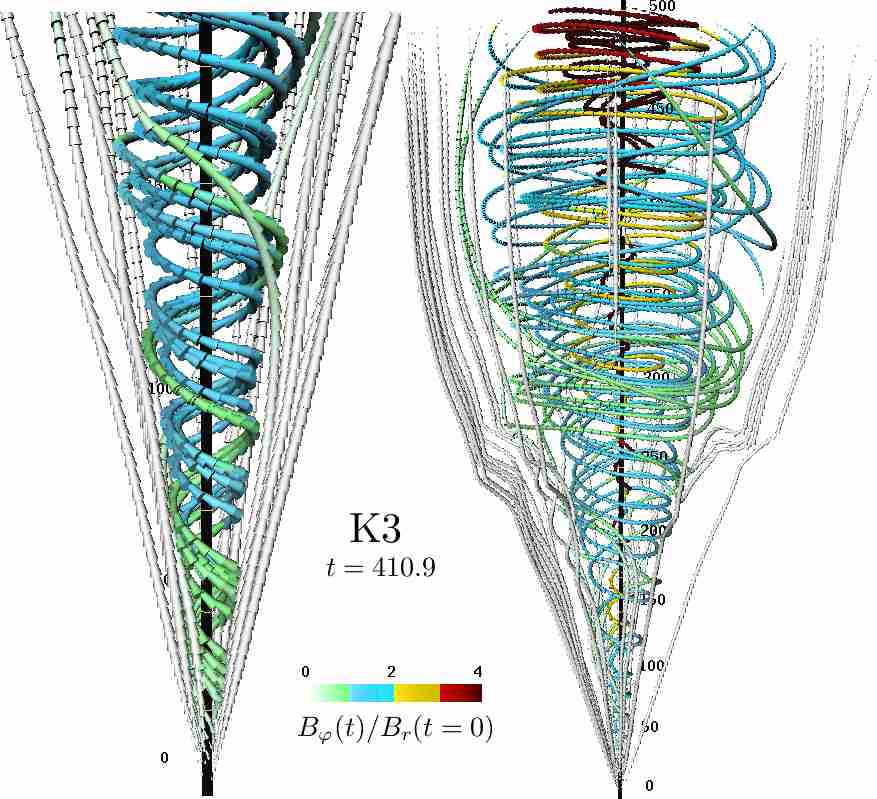}
\end{center}
\caption{Selected magnetic field lines in the 3D simulation with a Keplerian
velocity profile. The right-hand plot shows the entire domain ($r=5\ldots505$),
the left-hand plot only the lower part up to $r \approx 200$.  The color coding
represents the strength of the azimuthal magnetic field. The magnetic field
lines, which were purely radial to begin with, wind many times around the
central axis, rendering it susceptible to kink instabilities.}
\label{fig:flines}
\end{figure}

\begin{figure}[t]
\begin{center}
\includegraphics[width=.8\linewidth]{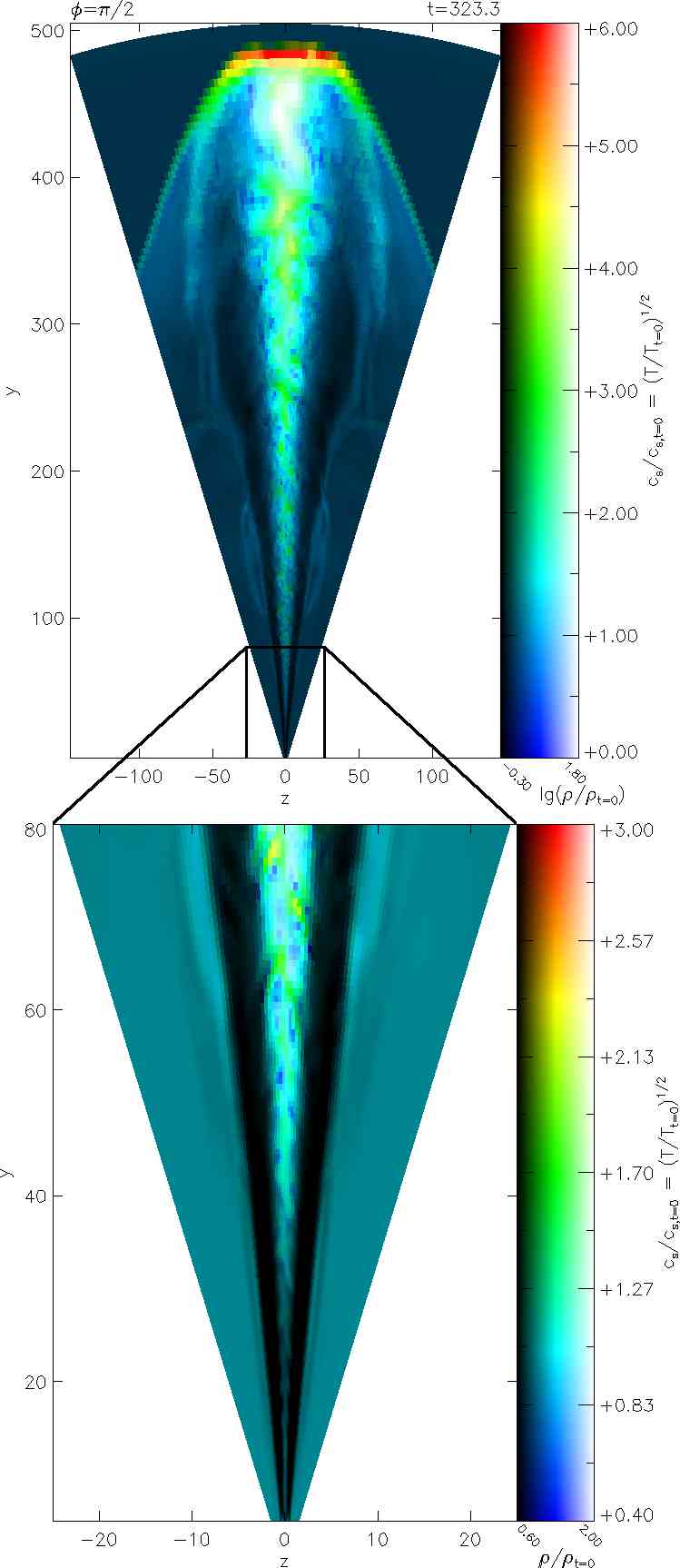}
\end{center}
\caption{Density and temperature in a meridional slice in simulation K3
(Keplerian rotation profile, 3D).  The density is encoded in intensity, with
bright colors representing regions that are overdense with respect to the
environment.  The (square root of the) temperature is encoded in the hue, with
blue meaning cold and red meaning hot with respect to the environment.  The
hoop stress associated with $B_\varphi$ squeezes the plasma towards the central
axis and creates an underdense cavity around the central part of the jet. At
the boundary of this cavity the environment exerts the stress that confines the
jet. The observed opening angle for a jet like this would be smaller than the
width of the cavity.}
\label{fig:rhot}
\end{figure}

The jets are initially accelerated mainly by gas pressure. This holds up to the
sonic surface, which lies about halfway to the \Alfven surface.  Then, the
Lorentz force becomes the dominant driving force.  The \Alfven radii are at $r
\approx 30\ldots140$, depending on the simulation and the direction
$\vartheta$: near the axis (small $\vartheta$), the poloidal field $B_r$ is
amplified and the \Alfven radii are at larger distances than in the outer
regions (large $\vartheta$), where $B_r$ is attenuated.  The opening angle of
the jets increase somewhat with distance, but to a first approximation the flow
can still be treated as conical, see Fig.~\ref{fig:border}.  The jet front
crosses the upper boundary ($r=505$) at $t\approx260$ in the 2.5D simulations
and at $t\approx330$ in the 3D simulations. The latter are subject to more
numerical dissipation of kinetic energy, because the grid there is not
symmetric in azimuthal direction. This reduces the injected Poynting flux, see
Fig.~\ref{fig:eflow}. Apart from that, 2.5D and 3D simulations give, for our
purposes, comparable results.  Fig.~\ref{fig:flines} shows how the magnetic
field is wound up inside the jet in one of the simulations.  A typical density
and temperature distribution is shown in Fig.~\ref{fig:rhot}.

\subsection{Expected instabilities}
\label{sec:liability}

\begin{figure}[t]
\begin{tabular}{@{}l@{}r}
\includegraphics[width=.5\linewidth]{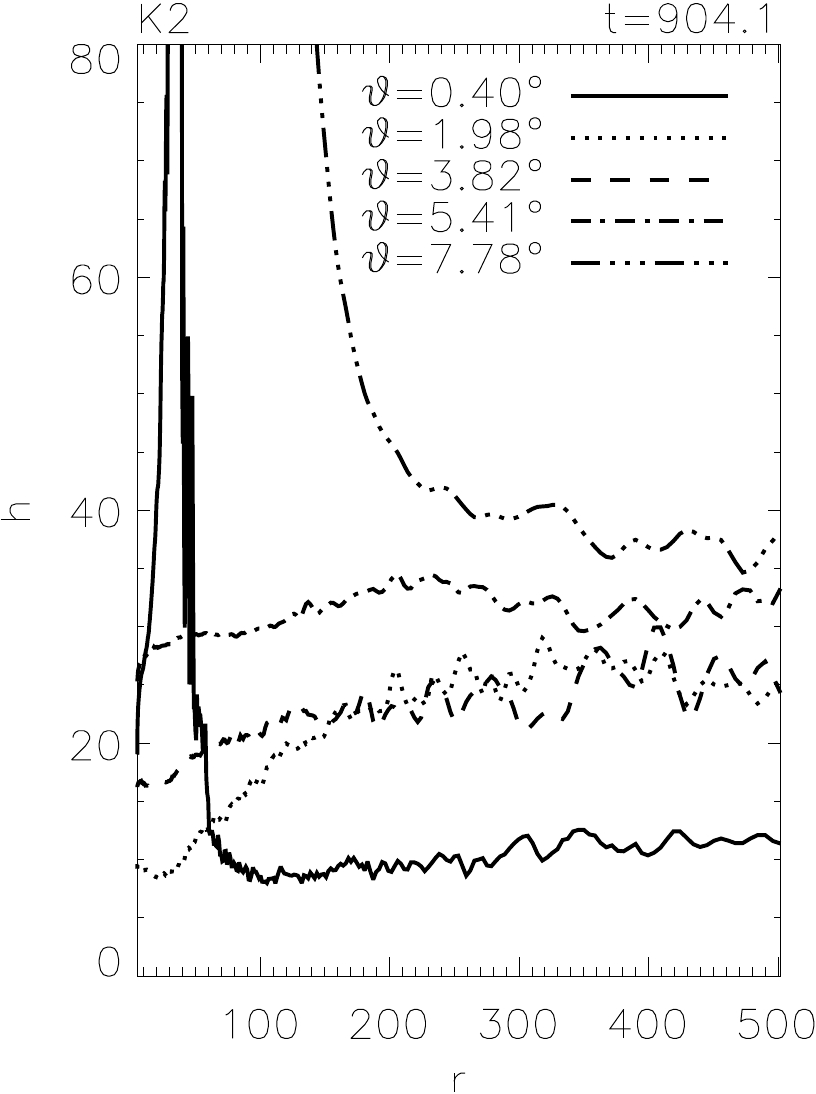} &
\includegraphics[width=.5\linewidth]{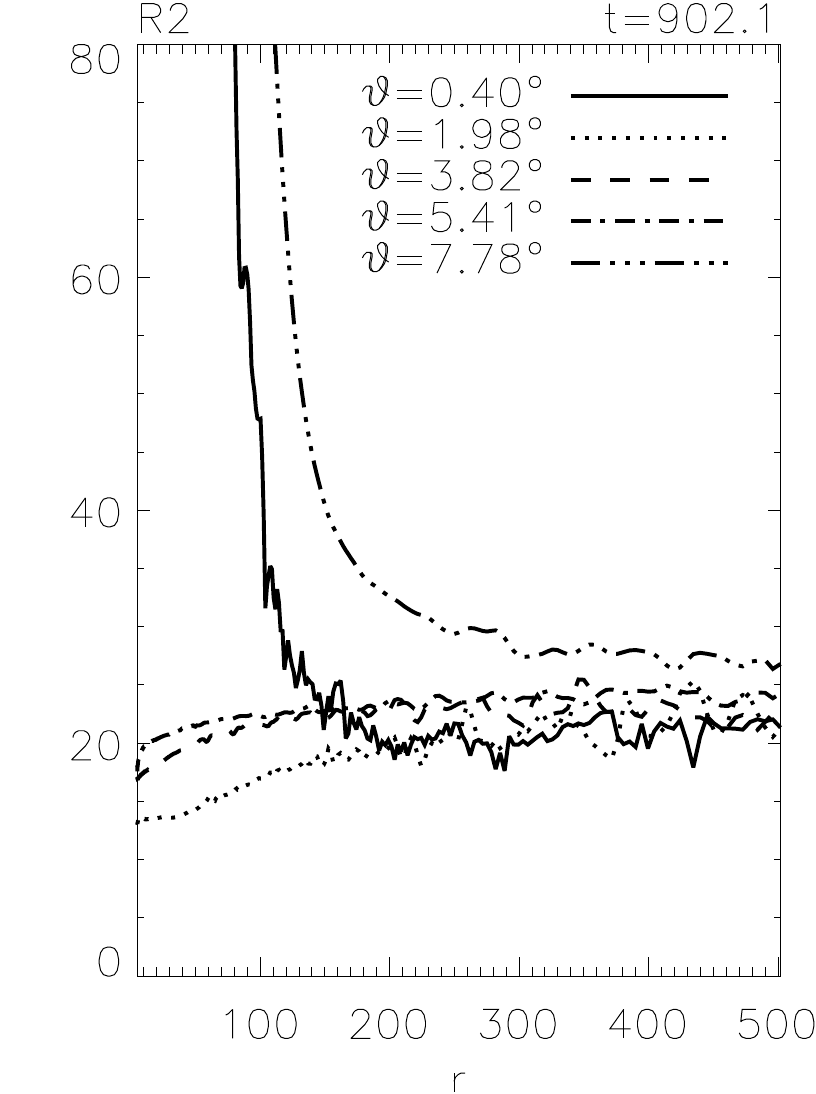}
\end{tabular}
\caption{Magnetic pitch as a function of distance in the 2.5D simulations. The
magnetic pitch is also the smallest possible wavelength of an instability. The
dependence of $h$ on the polar angle $\vartheta$ is stronger in the Keplerian
case (left plot), suggesting higher stability.}
\label{fig:pitch}
\end{figure}

\begin{figure}[t]
\begin{tabular}{@{}l@{}r}
\includegraphics[width=.5\linewidth]{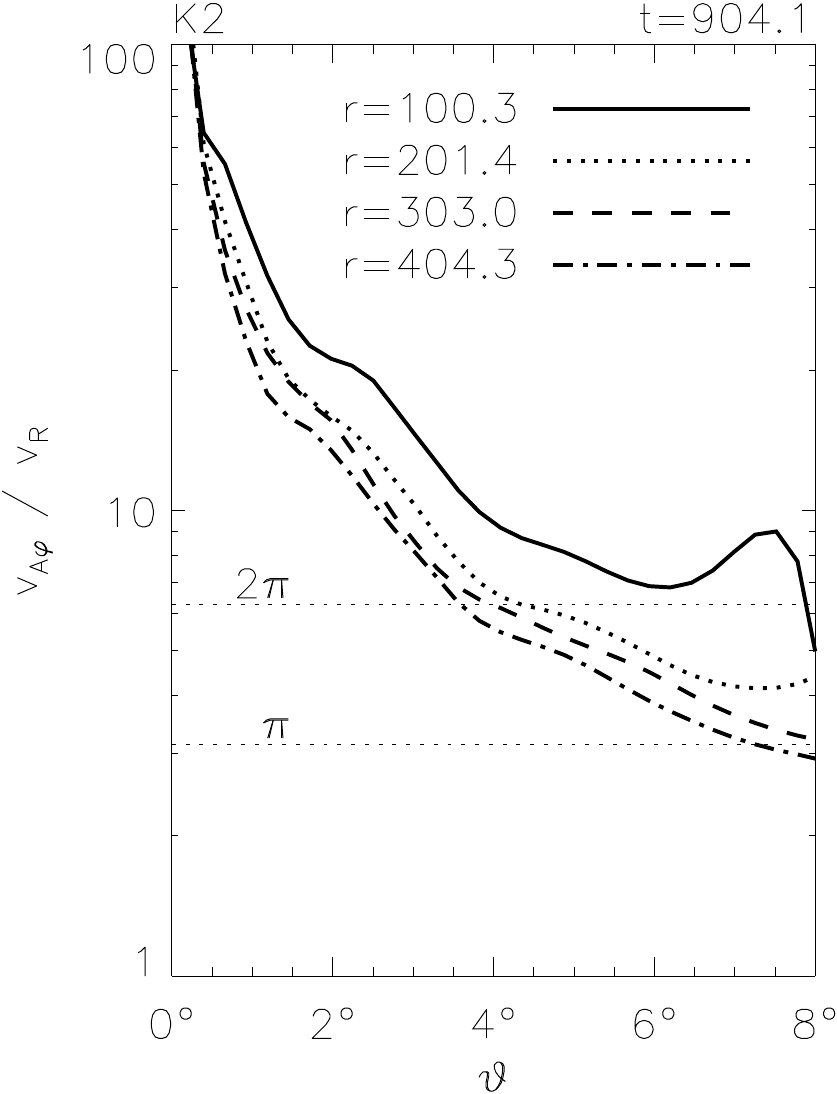} &
\includegraphics[width=.5\linewidth]{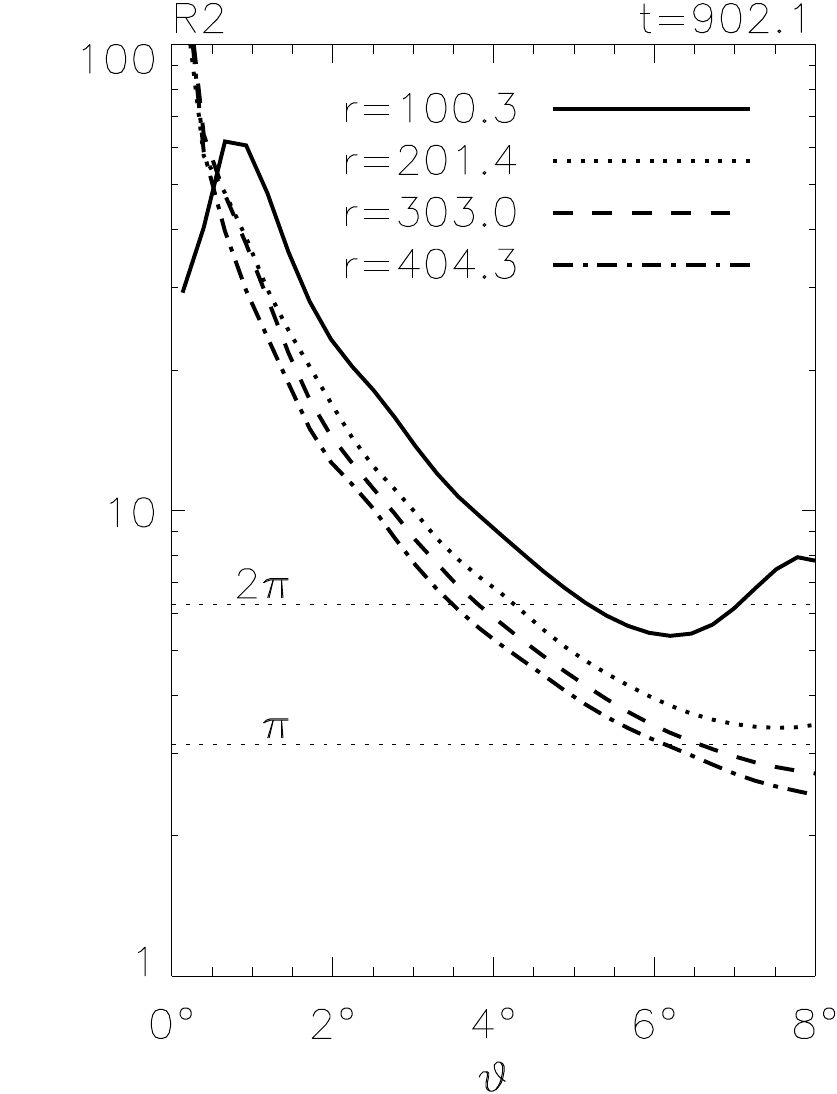}
\end{tabular}
\caption{The expected onset of instability depends on the ratio of the
azimuthal \Alfven speed and the expansion velocity. This ratio is shown here
for the 2.5D simulations. The horizontal dotted lines are for two estimates of
the condition under which growth is possible (see text). Below the respective
line, expansion prevails and an instability cannot grow.}
\label{fig:growthcondition}
\end{figure}

\begin{figure}[t]
\begin{tabular}{@{}l@{}r}
\includegraphics[width=.5\linewidth]{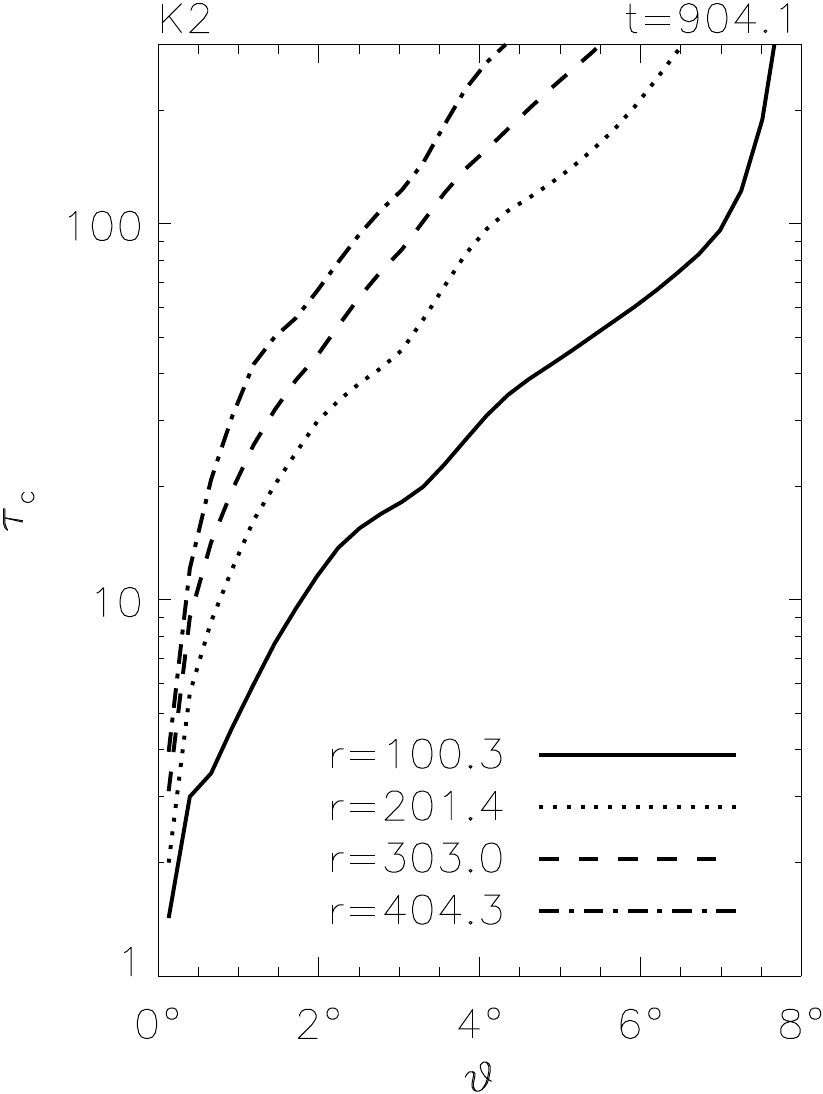} &
\includegraphics[width=.5\linewidth]{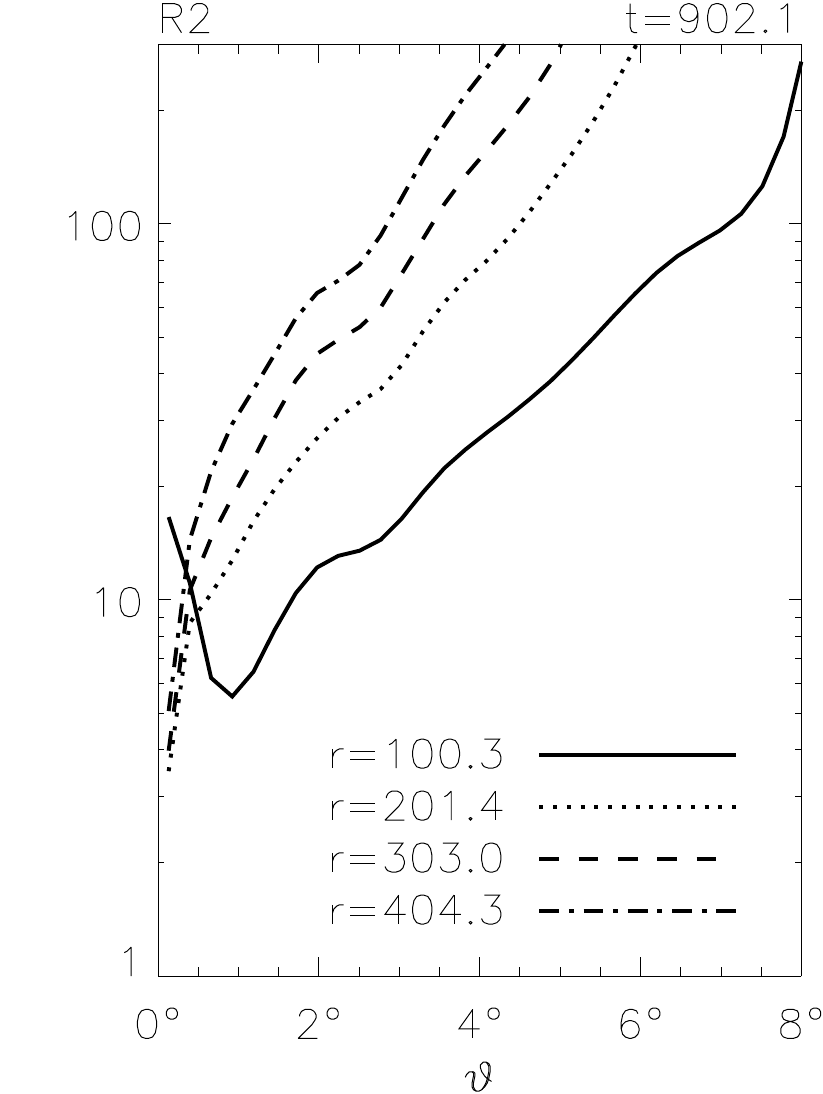}
\end{tabular}
\caption{\Alfven crossing times with $\iota = \pi$ in the 2.5D simulations.  In
both cases, $\tauc$ and with it the expected instability growth time increases
with radius and distance from the central axis. The curves for $r \approx 100$
represent an underestimate, because the jet still accelerates at this
distance.}
\label{fig:acrosstime}
\end{figure}

The observed instabilities can be compared with expectations from linear
stability theory. To do this, we extract from the axisymmetric simulations the
quantities  that enter the stability conditions, and then compare the result
with the evidence of non-axisymmetric instability in the corresponding 3D
simulation. The available stability conditions apply only to steady or static
configurations and have been derived either in the context of controlled fusion
or cylindrical jets \citep{2000Appl,2000Lery}, hence the comparison can only be
indicative.

According to the Kruskal-Shafranov criterion, the longitudinal wavelength of an
instability must be at least as high as the magnetic pitch on the unstable
surface, defined to be the distance covered during one revolution of a helical
field line about the central axis.  Besides being the result of linear
stability analyses in the context of controlled fusion, it can be derived
heuristically from geometric arguments \citep{1958Johnson}.  Therefore, it
should give a convenient scale also in cases for which it was not originally
intended, like the expanding jets studied here.  Deviations can be expected
e.g. from the effect of one-ended line-tying, which in some cases has been
found to lead to increased instability as opposed to a configuration without a
free end \citep{2006Furno,2006Lapenta,2008Sun}.

For a conical jet, the magnetic pitch is
\begin{equation}
    h = 2\pi R \abs{\frac{B_r}{B_\varphi}}
\label{eq:pitch}
\end{equation}
on the $\vartheta=\const$ surface. See Appendix~\ref{app:conicalpitch} for a
derivation of this expression. In the simulations, $h$ decreases with $r$ and
settles to a constant value above the \Alfven radius, see Fig.~\ref{fig:pitch}
(compare also Fig.~\ref{fig:flines}). The variation of the pitch with
$\vartheta$ depends on the kind of rotation imposed at the lower boundary. In
the Keplerian case, the dependence is strong, with the asymptotic pitch being
approximately $10$ near the axis and $40$ at the limb of the jet.  In the rigid
rotation case, $h \approx 25$ in all directions within the jet.

The \Alfven crossing time in a conical, unaccelerated jet, defined as the time
it takes an azimuthal \Alfven wave to orbit the central axis, is given by
\begin{equation} \tauc = \frac{\iota R}{\vAphi - \iota v_R}
\label{eq:acrosstime} \end{equation} where $\iota=2\pi$ for a full revolution,
$\vAphi=\const$ is the azimuthal \Alfven speed and $v_R = v_r \sin \vartheta$
is the expansion velocity.  In the simulations, $\vAphi \approx \const$ above
the \Alfven radius.  This is as expected theoretically from conservation of
mass and magnetic flux in a conically expanding, steady axisymmetric jet.
$\tauc$ is finite and physically meaningful only if the condition
\begin{equation}
    \frac{\vAphi}{v_R} > \iota
\label{eq:growthcondition}
\end{equation}
is satisfied. If it is not, the expansion takes place too fast for an \Alfven
wave to cross the jet and an \Alfven mode instability cannot grow.  While the
critical value of $\iota$ is arguable, we note that causal contact across the
jet by Alfv\'en waves is only possible if $\iota \ge \pi$.  The situation in
our simulations is illustrated in Figs.~\ref{fig:growthcondition}
and~\ref{fig:acrosstime}. Instabilities can grow only slowly on magnetic
surfaces with large $\vartheta$.  Depending on the $\iota$ needed for efficient
growth, they may even be stalled due to the jet's expansion.  In any case,
instabilities grow most rapidly if they start at small $r$.  In regions where
the jet is accelerating ($\de v_r / \de r > 0$) or decollimating ($\de
\vartheta / \de r > 0$ along a field line), the effective \Alfven crossing time
is underestimated by Eq.~\eqref{eq:acrosstime}. The jet is then stabler than
condition~\eqref{eq:growthcondition} suggests.

\subsection{Instabilities found in the simulation}

\begin{figure}[t]
\includegraphics[width=\linewidth]{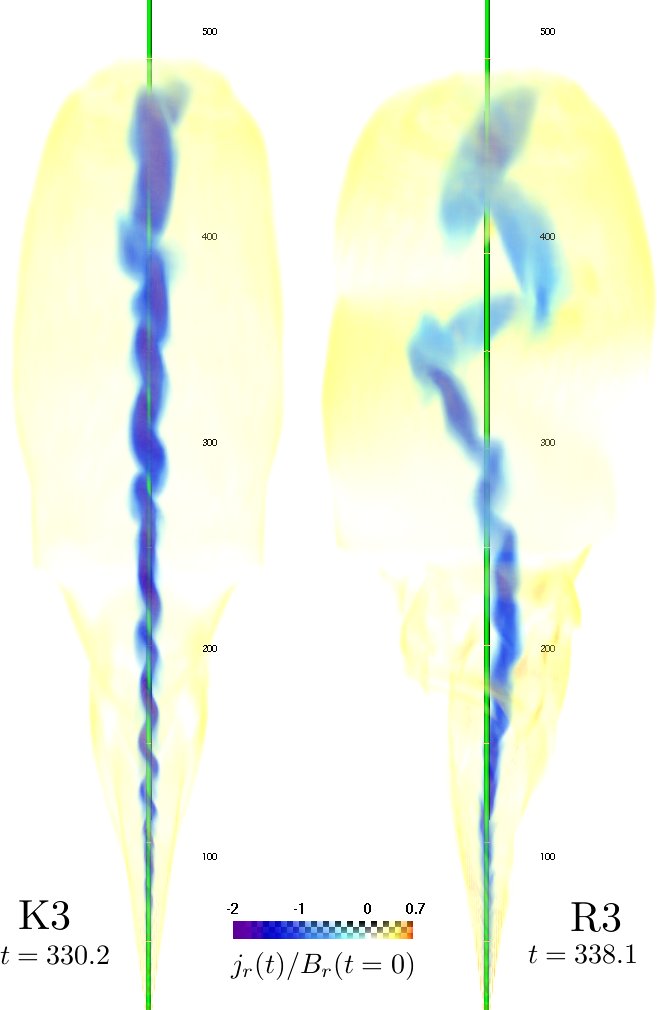}
\caption{Radial component of the current density ($\Nabla \times \vec{B}$) in
the 3D simulations just before the jets reach the upper boundary. The rod in
the middle of the jets is their central axis ($y$-axis).  Helical distortions,
characteristical for kink instabilities, can be seen in the backward current
(blue) in both cases.  The amplitudes and wavelengths are significantly larger
in the simulation with a rigid rotation profile (right-hand image).}
\label{fig:jrvolren}
\end{figure}

\begin{figure}[t]
\includegraphics[width=\linewidth]{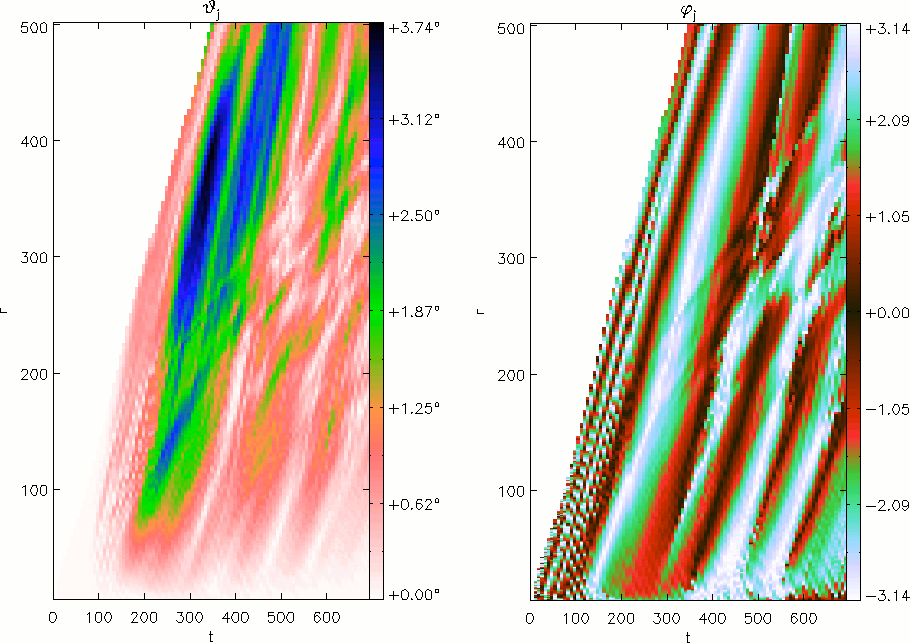}
\caption{Position of the barycenter of the backward current (blue material in
Fig.~\ref{fig:jrvolren}) in simulation R3. The left-hand map shows the
amplitude of the instabilities and the right-hand one its phase. The blank
region is where the jet has not been yet, its border marks the jet front.
Observers moving with the flow follow a time-position curve which is
(approximately) parallel to the front, i.e.  the instabilities are at rest with
respect to a comoving frame.}
\label{fig:jrmbarycenter}
\end{figure}

\begin{figure}[t]
\includegraphics[clip=true,width=\linewidth]{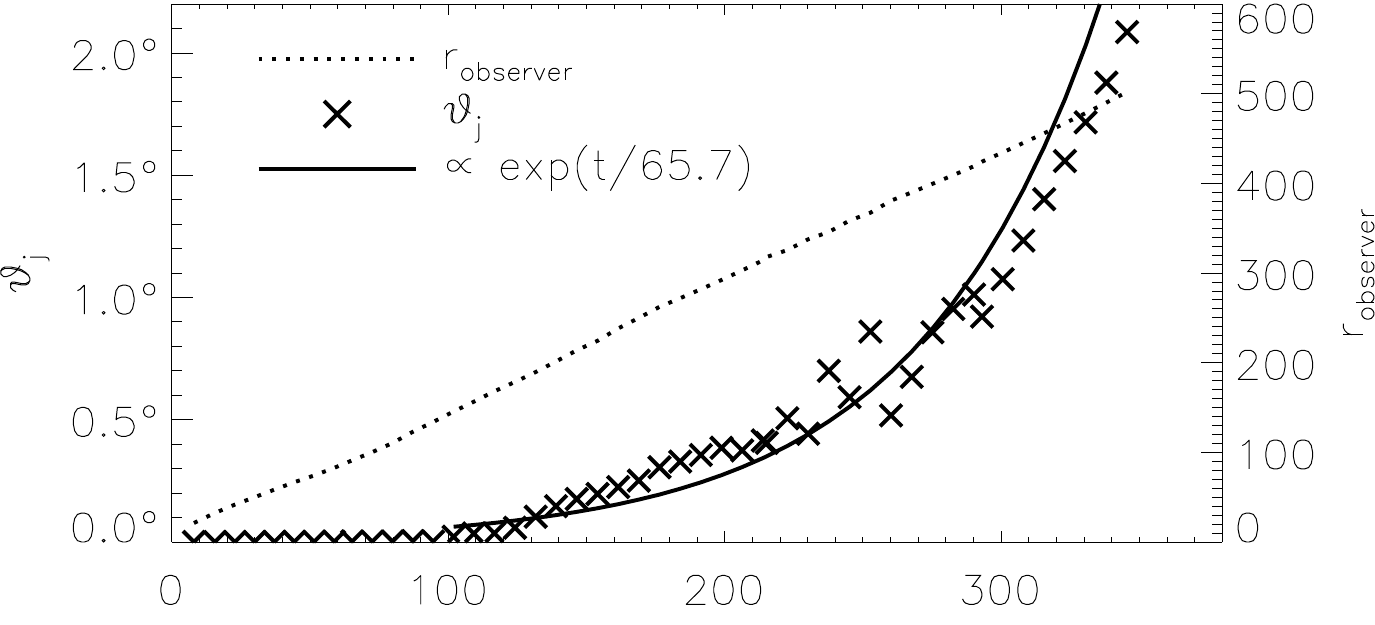} \\
\includegraphics[width=\linewidth]{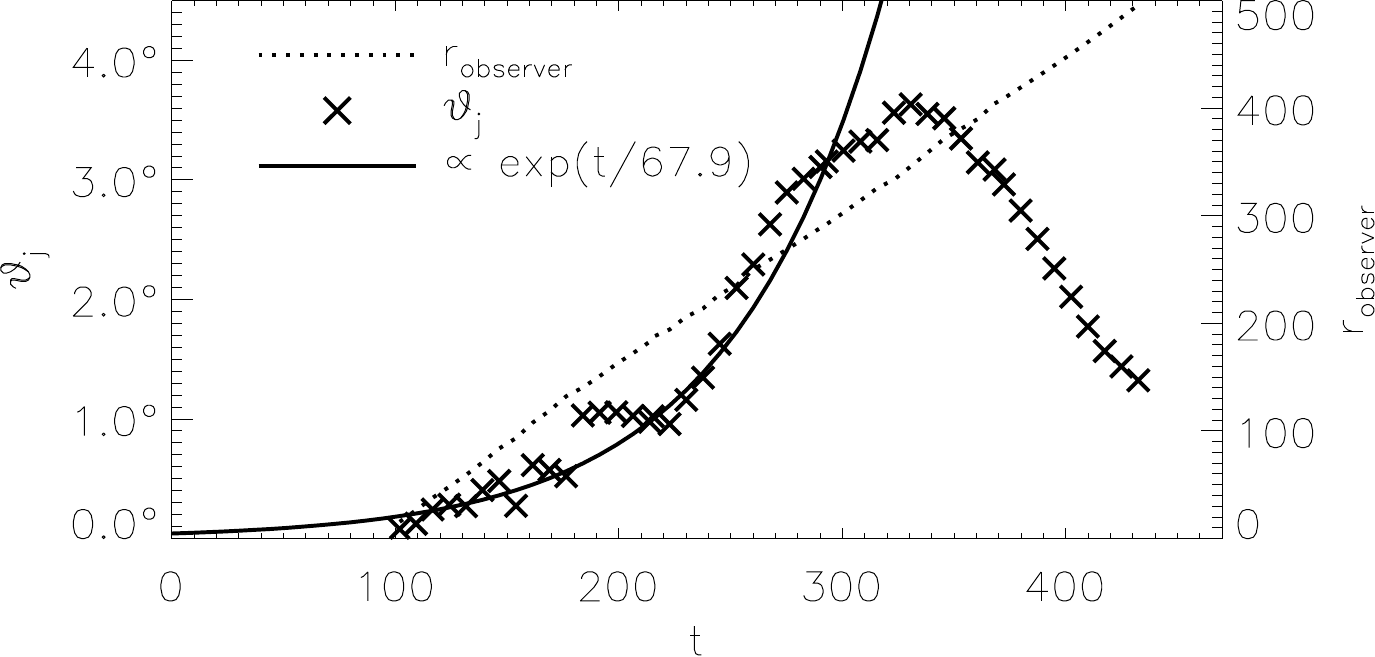}
\caption{Amplitude of the perturbations (crosses, left-hand axis) as seen by an
observer (dotted line, right-hand axis) located just behind the jet front
(upper panel) and located 130 length units behind the jet front (lower panel)
in simulation R3.  The exponential fits (solid line, left-hand axis) yield the
instability growth time.}
\label{fig:comovingthj}
\end{figure}

We observe non-axisymmetric, kink-like distortions in the magnetic field and
other quantities in both 3D simulations. They emerge near the \Alfven radius,
propagate with the flow and grow in amplitude along the way.  It is convenient
to look at the current density $\vec{j}=\frac{c}{4\pi}\Nabla \times \vec{B}$
for a quantitative analysis.  The radial component $j_r$ is related to
$B_\varphi$ and is as such characteristic for the distortions in the magnetic
field.  In the unperturbed case, it is concentrated about the central axis and
along the outer boundary of the cavity illustrated in Fig.~\ref{fig:rhot}, with
respectively opposite orientation.  In our simulations, $B_\varphi$ is directed
in negative $\varphi$-direction and the axial current, accordingly, in negative
$r$-direction. We denote this backward current with $j_r^-$, so that $j_r =
j_r^+ + j_r^-$.

In the rigid rotation case (R3), the distortions attained large amplitudes of
several degrees. Looking at $B_\varphi$ in the $r=\const$ plane, we find that
the whole jet is affected by the kink.  The number of visible radial nodes is
2--4, corresponding to wavelengths on the order of $150$, i.e. several times
larger than the magnetic pitch.  Owing to the distortions in the magnetic
field, the axial current was perturbed as shown in Fig.~\ref{fig:jrvolren} on
the right-hand image.  $j_r^-$ helically twines around the central axis in
reminiscence of ``ideal'' kink instabilities with an azimuthal mode number
$m=1$.

To analyze the unstable displacements, we determine the barycenter of the
backward current $j_r^-$ in the $r=\const$ plane, denoting its location with
$(\vartheta_j,\varphi_j)$. The result, from which one can directly read off
amplitudes and wavelengths, is shown in Fig.~\ref{fig:jrmbarycenter}.  The
slope of the points of constant phase $\varphi_j$ in the $r$-$t$ diagram
corresponds with the flow velocity $v_r$.  That is, the instabilities are at
rest with respect to a comoving frame.  We estimate the growth time in such a
frame by introducing an observer moving with flow and measure $\vartheta_j$ in
doing so.  We find strictly increasing, exponential growth if the observer is
located just behind the jet front, see upper panel in
Fig.~\ref{fig:comovingthj}. The exponential growth time $\taug$  is generally
on the order of the \Alfven crossing times shown in Fig.~\ref{fig:acrosstime}.
For observers which are farther behind the jet front, the amplitude does not
follow a simple exponential increase, see lower panel in
Fig.~\ref{fig:comovingthj} for an example. Rather, it saturates and even
declines in some cases. The reason for this is not clear.  We cannot rule out
the possibility that there is stabilizing feedback from the upper boundary.
Considering that the flow is super-\Alfvenic there, this seems unlikely
though.

In the Keplerian rotation case (K3), the jet also exhibits kink-like
distortions, see left-hand image in Fig.~\ref{fig:jrvolren}. However, the
perturbation amplitudes are much smaller, with $\vartheta_j$ attaining peak
values of about $1.4\degree$ directly behind the jet front and only about
$0.5\degree$ farther behind.  Unlike in the rigid rotation case, only inner
regions of the jet are affected by the kinks, the jet border is relatively
unharmed.  The wavelengths are on the order of 25--50, i.e. there are more
radial nodes than in the rigid rotation case.  Even for an observer traveling
just behind the jet front, the amplitude is not strictly increasing, but
saturates and tapers off after an initial rise with a growth time of $\taug
\approx 50$, i.e. also on the order of the relevant \Alfven crossing times in
Fig.~\ref{fig:acrosstime}.

\subsection{Impact on dynamics and energetics}

\begin{figure}[t]
\includegraphics[width=\linewidth]{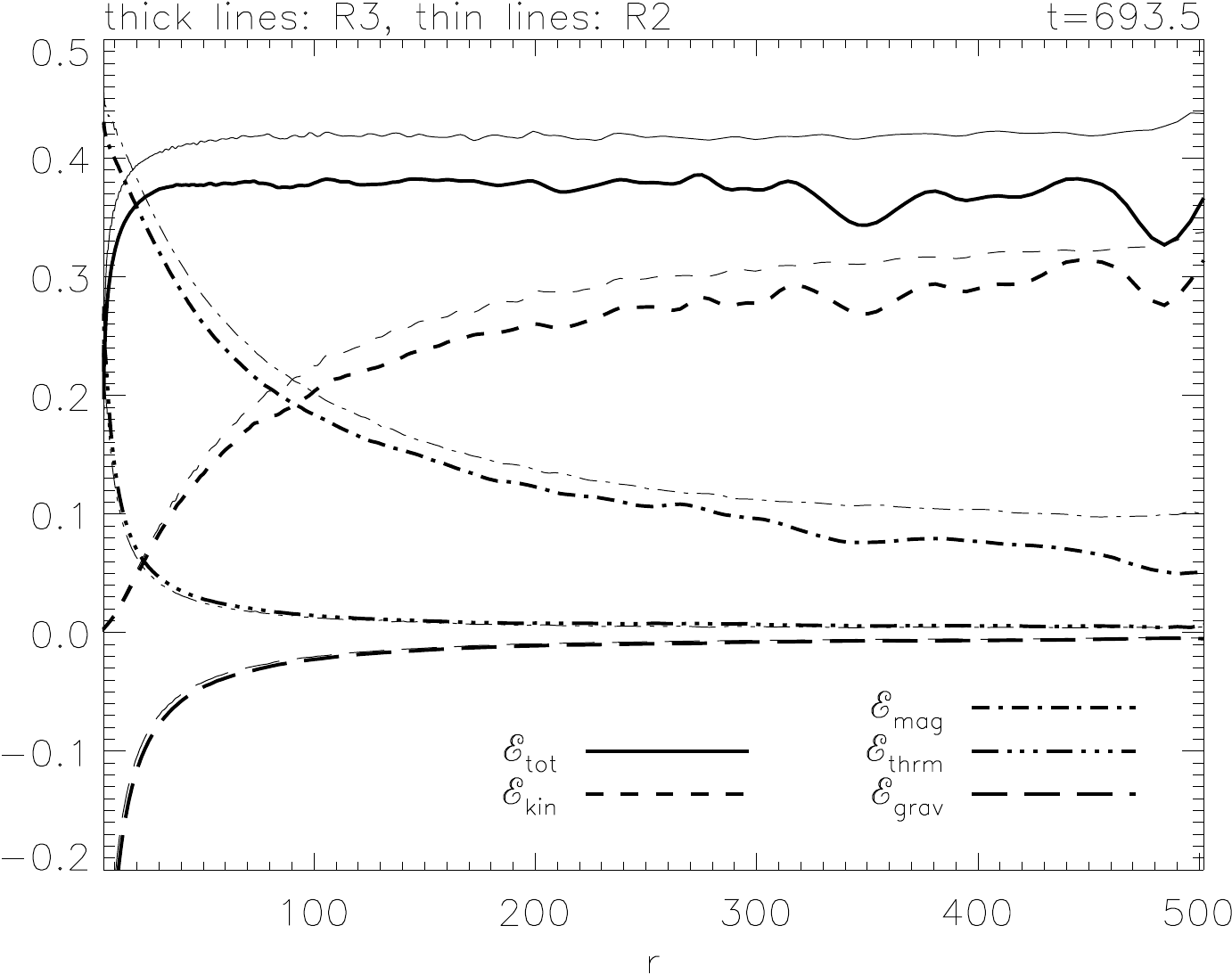}
\caption{Energy flow rates through $r=\const$ in 3D (thick lines) and in the
2.5D (thin lines) simulations with a rigid rotation profile. There is no clear
evidence of additional conversion of Poynting flux to kinetic energy.  The
energy flow through the lateral boundaries is virtually zero at all times. The
situation is similar in the Keplerian case (K3 and K2).}
\label{fig:eflow}
\end{figure}

\begin{figure}[t]
\includegraphics[clip=true,width=\linewidth]{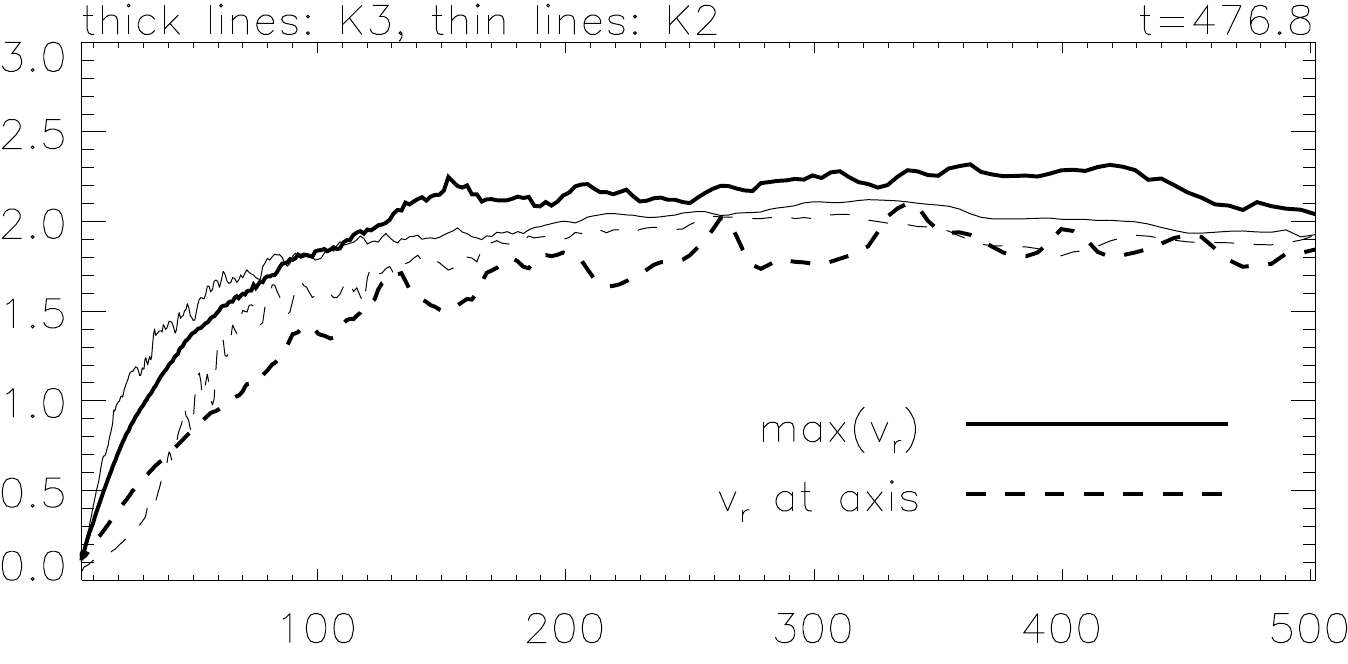} \\
\includegraphics[width=\linewidth]{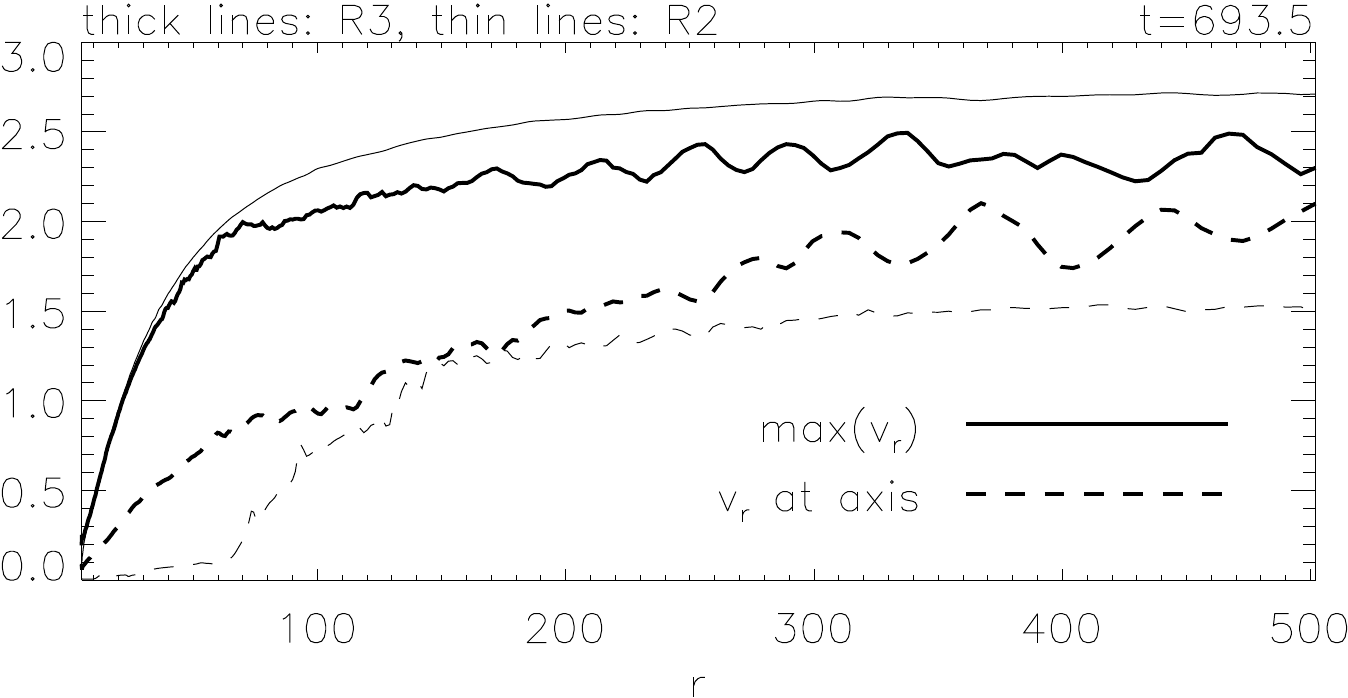}
\caption{Radial velocities in the 3D (thick lines) and in the 2.5D (thin lines)
simulations.  The jets do not accelerate above the \Alfven radius, despite
instabilities. The value of $v_r$ near the central axis in simulation R3,
though rising, does not give conclusive evidence of acceleration, because the
jet's axis is strongly distorted by the instabilities.}
\label{fig:vr}
\end{figure}

Instabilities release magnetic energy by transforming it into kinetic energy, but
for dissipation in the sense of magnetic reconnection sufficiently small length
scales have to develop. In ideal MHD simulations like ours, such dissipation is
present in the form of numerical diffusion due to the effect of interpolations
across adjacent grid cells. The effect cannot be modeled by Ohmic resistivity
but can be quantified through secondary effects like changes in the energy
fluxes.

We did not find conclusive evidence of magnetic field dissipation provoked by
the instabilities in the 3D simulations.  For example, the flow of nonradial
magnetic field
\begin{equation}
    \int_{r=\const} \sqrt{B_\vartheta^2+B_\varphi^2}\,  v_r \,\de A 
\end{equation}
shows asymptotically the expected $\simm r$ behavior (as $B_\varphi \sim
r^{-1}$ and $A \sim r^2$) with minor fluctuations but no decreasing trend when
compared to the 2.5D simulations.

It is helpful to look at the energy fluxes.  From
Eq.~\eqref{eq:energy_equation_alt}, the energy flow rate in poloidal direction
is
\begin{equation}
    \efrate_\text{tot}(t,r) = \int_{r=\const} \left[ \left( 
            \frac{1}{2} \rho v^2 + \frac{\gamma}{\gamma-1}p + \rho \Phi \right) v_r
            + S_r \right] \de A .
\label{eq:efrate}
\end{equation}
Decomposing the integral from left to right, we identify the flow rates of
kinetic energy $\efrate_\text{kin}$, thermal enthalpy $\efrate_\text{thrm}$,
gravitational potential energy $\efrate_\text{grav}$ and magnetic enthalpy
$\efrate_\text{mag}$.  We plotted these in Fig.~\ref{fig:eflow} for the
simulations R3 and R2.  The total energy flow $\efrate_\text{tot}$ rises for
small $r$ due to the temperature-control term $K$ in
Eq.~\eqref{eq:energy_equation}.  The conversion of Poynting flux to kinetic
energy flux in the 3D simulation looks qualitatively the same as in the 2.5D
comparison simulation. In particular, there is no evidence of an additional
conversion of $\efrate_\text{mag}$ to $\efrate_\text{kin}$ due to magnetic
field dissipation. This agrees with the fact that there is no additional
acceleration of the jets, see Fig.~\ref{fig:vr}.

\section{Discussion and conclusions}
\label{sec:discussion}

We have simulated magnetocentrifugally driven, conical jets over a range in
distance of 1000 times the initial jet radius, in both 3D and axisymmetric
2.5D.  The calculations extend to a factor of about 5--10 beyond the \Alfven
surface.  The 3D jets developed non-axisymmetric instabilities of the kink
kind.

The violence of the instabilities depends on the rotation profile applied at
the base.  With a rigid rotation ($\proptoo R$) profile, the perturbations grow
to much larger amplitudes than with a Keplerian ($\proptoo R^{-1/2}$) profile.
We suspect that the reason for the differing behavior lies in the magnetic
shear, defined as the variation of the magnetic pitch with distance to the
axis. In the rigid rotation case, there is virtually no shear as opposed to the
Keplerian case, for which the pitch increases with distance from the axis,
see Fig.~\ref{fig:pitch}.  A shear-free configuration is expected to be
unstable to non-resonant modes, whereas a configuration with increasing pitch
is expected to be unstable to modes with a resonant surface inside the jet
\citep{2000Appl,2000Lery}.  This fits well with what we observe in the
simulations, viz. that the kink is confined inside the jet in the Keplerian
case. Heuristically speaking, the differing behavior could be attributed to the
fact that the outer (high $\vartheta$) layers of the jet, which are more stable
(higher magnetic pitch), damp internally arising modes in the Keplerian case.

In both cases, the longitudinal wavelength of the instabilities is $\simm5$
times larger than the value of the magnetic pitch near the axis. The relation
is qualitatively consistent with the findings of \citet{2000Appl} for a
cylindrical jet. The growth time of the instabilities is on the order of the
\Alfven crossing time.  The exact relation is difficult to determine, because
the crossing time as well as the location of the resonant surface can only be
estimated.  As the azimuthal magnetic field strength and with it the azimuthal
\Alfven speed decrease past the \Alfven surface, opposing parts of the jets
become causally disconnected from each other. Thus, the jet expands too fast
for \Alfven mode instabilities to grow. The effect is amplified if the jet is
diverging.  Recollimation, on the other hand, should boost the growth of
instabilities.

As found in other studies, the conversion of magnetic enthalpy (Poynting flux)
to kinetic energy is fairly efficient, on the order 70\%. Dissipation of
magnetic fields by internal instabilities is expected to contribute additional
acceleration of the flow \citep{2002Drenkhahn}.  The calculations do not show a
clear signature of this process.  It seems that either the observed region is
too small, and/or the numerical dissipation of magnetic fields is too weak.
Also, from a macroscopic point of view, the instabilities were not violent
enough to bring together fields with an antiparallel component, as is necessary
for magnetic reconnection to occur.  Moreover, most of the magnetic enthalpy
was already converted in the magnetocentrifugal acceleration process.
Therefore, even if there was magnetic dissipation, the effect would not be
dramatic. Nevertheless, we found that the magnetic field gets significantly
distorted by the instabilities.  This should facilitate magnetic field
dissipation further downstream but it may be necessary to extend the
calculations to larger distances to see the effect.

It is tempting to compare the instability-related structures in the simulations
with structures in observed jets. The 3D jet structure in Fig.~\ref{fig:rhot},
for example, is reminiscent of the semi-regular patterns seen in \Halpha images
taken of outflows from young stellar objects (YSO) like HH 34
\citep{2002Reipurth}.  There are, however, other possible interpretations of
the observed structure. The wiggles in YSO jets could also be the result of a
precessing or orbitally moving source \citep[and references
therein]{2002Masciadri}. The symmetric nature of structures often seen in jet
and counterjet \citep[e.g. HH 212][]{2001Wiseman} suggests a modulation of the
outflow speed or mass flux originating at the source of the outflow rather than
an instability developing further away. The irregularities caused by the
instabilities studied here are possibly more important for internal magnetic
energy release inside the jet than for major observable structures like the
knots and wiggles in YSO jets, though they are likely to contribute to these at
some level as well.

\begin{acknowledgements}
We thank an anonymous referee for constructive comments. We are also grateful
to Dimitrios Giannios for stimulating discussions.
\end{acknowledgements}

\bibliography{ref}

\appendix

\section{Magnetic pitch for a conical jet}

\label{app:conicalpitch}
\begin{figure}[t]
\begin{center}
\includegraphics{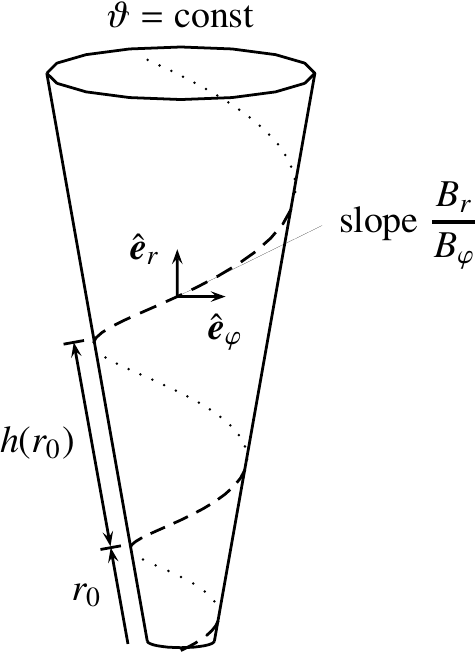}
\end{center}
\caption{Magnetic field line with pitch $h$ on a conical surface.}
\label{fig:conicalpitch}
\end{figure}

The radial progress of the field line depicted in Fig.~\ref{fig:conicalpitch}
is determined by
\begin{equation}
    \frac{\de r}{R \de \varphi} = \abs{\frac{B_r}{B_\varphi}} \eqqcolon a(r)
\end{equation}
where $R=r \sin\vartheta$ is the distance to the polar axis and $a(r)$ is
the unsigned slope. Assuming that $B_r \propto r^{-2}$ and $B_\varphi \propto
r^{-1}$ due to magnetic flux conservation, we can write
\begin{equation}
    a(r) = a_0 \frac{r_0}{r} .
\end{equation}
By integrating the resulting expression we obtain the radial distance covered
after one revolution:
\begin{equation}
    \int_{r_0}^{r_0+h} \de r = \int_0^{2\pi} a_0 R_0 \, \de\varphi \quad \Rightarrow \quad h = 2\pi a_0 R_0 .
\end{equation}
Alternatively, we could also define a local magnetic pitch $\tilde{h}$ by
taking $a = a_0 = \const$ for the slope. The result is
\begin{equation}
    \frac{\tilde{h}}{r_0} = \exp\left( 2\pi a_0 \sin\vartheta \right) - 1 .
\end{equation}
The difference between $h$ and $\tilde{h}$ turned out to be insignificant in
our analysis.  This is understandable since $\tilde{h} \rightarrow h$ for small
$a_0 \sin\vartheta$.

\end{document}